%% file: main.tex
\documentclass[conference]{IEEEtran}
\usepackage{cite}
\usepackage{amsmath,amssymb,amsfonts}
\usepackage{algorithmic}
\usepackage{graphicx}
\usepackage{textcomp}
\usepackage{xcolor}
\usepackage[hyphens]{url}

\def\BibTeX{{\rm B\kern-.05em{\sc i\kern-.025em b}\kern-.08em
    T\kern-.1667em\lower.7ex\hbox{E}\kern-.125emX}}

\usepackage{tikz}
\usepackage{subfigure}
\usepackage{color}
\usepackage{pifont}
\usepackage{enumitem}
\usepackage{multirow}
\usepackage{diagbox}

\newcommand{\tabincell}[2]{\begin{tabular}{@{}#1@{}}#2\end{tabular}}
\newcommand*{\circled}[1]{\lower.7ex\hbox{\tikz\draw (0pt, 0pt)%
		circle (.5em) node {\makebox[1em][c]{\small #1}};}}
\newcommand{\secref}[1]{\S\ref{#1}} 

\title{From Ideal to Practice: Data Encryption in eADR-based Secure Non-Volatile Memory Systems}
\author{
	{\rm Jianming Huang, Yu Hua}
	\\
	Huazhong University of Science and Technology \\
}

\begin{document}
\maketitle
\thispagestyle{plain}
\pagestyle{plain}


\begin{abstract}
Extended Asynchronous DRAM Refresh (eADR) proposed by Intel extends the persistence domain from the Non-Volatile Memory (NVM) to CPU caches and offers the persistence guarantee. Due to allowing lazy persistence and decreasing the amounts of instructions, eADR-based NVM systems significantly improve performance. Existing designs however fail to provide efficient encryption schemes to ensure data confidentiality in eADR-based NVM systems. It is challenging to guarantee both data persistence and confidentiality in a cost-efficient manner due to the transient persistence property of caches in eADR. Once the system crashes, eADR flushes the unencrypted data from the cache into NVM, in which security issues occur due to no encryption. To bridge the gap between persistence and confidentiality, we propose cost-efficient BBE and Sepencr\footnote{This paper is an extended version of our published paper in IEEE Computer Architecture Letters (DOI: 10.1109/LCA.2022.3225949)} encryption schemes that efficiently match different eADR execution models from ideal to practice. Under the ideal eADR execution model, BBE supports the encryption module via the battery of eADR upon crashes. Under the practical eADR execution model, Sepencr generates the one-time paddings (OTPs) at the system startup to encrypt the cached data in case the system crashes. Our evaluation results show that compared with an intuitive in-cache encryption scheme in eADR-based systems, our designs significantly reduce performance overheads while efficiently ensuring data confidentiality.

\end{abstract}

\input{introduction-new.tex}

\input{background.tex}

\input{design.tex}
\input{experiment.tex}

\input{discussion.tex}

\input{related.tex}
\input{conclusion.tex}


\bibliographystyle{IEEEtranS}
\bibliography{references}

\end{document}

%% file: introduction-new.tex
\section{Introduction}
\label{section 1}


Non-Volatile Memory (NVM) provides near-DRAM performance, low standby power consumption, and disk-like durability~\cite{hassan2015software,YueZ13,izraelevitz2019basic}. The byte-addressable feature of NVM is efficient to deliver high performance for NVM devices. NVM, as the Persistent Memory (PM)~\cite{yihtmfs}, allows the persistence boundary to move from storages, e.g., disk and SSD, to the memory~\cite{kadekodi2019splitfs,xu2016nova,xu2017nova,yihtmfs}. Moreover, Intel proposes Asynchronous DRAM Refresh (ADR)~\cite{ADR} and extended ADR (eADR)~\cite{eADR} to further extend the persistence domain to the on-chip memory controller (MC) and CPU caches. Specifically, ADR guarantees that the write pending queue (WPQ) in the memory controller becomes a persistent domain by flushing the data from WPQ into NVM upon power-down via the backup battery. eADR further guarantees that all on-chip data buffers, including CPU caches, become persistent domains~\cite{alshboul2021bbb} by flushing data from these buffers into NVM upon crashes. By using ADR and eADR, the persistent domains are extended to the on-chip buffers.


The significant persistence extensions are efficient to deliver high performance, which however overlook the encryption schemes in the heterogeneous memory/storage devices. Data in the persistent domains need to be encrypted to ensure confidentiality, i.e., the data cannot be accessed by attackers. In existing storage architecture, different persistent devices leverage different encryption schemes. For the external storage, e.g., disk and SSD, the data-at-rest encryption schemes are used to encrypt data~\cite{dickens2018strongbox,casey2008impact}, e.g., full disk encryption (FDE) in the disk~\cite{khati2017full}. For the NVM, e.g., Intel Optane PM~\cite{IntelDCPM}, the standard 256-AES hardware encryption~\cite{PMEncryption} is used to encrypt data.


Since the on-chip memory controller (specifically, the WPQ in the memory controller) exists in the persistent domain due to the support of ADR, the memory controller also leverages an encryption scheme for data confidentiality. Unlike the disk and PM in the off-chip domain, the encryption/decryption latencies for the data in the on-chip memory controller significantly impact the system performance. To efficiently encrypt the data in the on-chip memory controller for data confidentiality, a low-overhead counter mode encryption (CME) scheme in the memory controller has been proposed and widely used in existing works~\cite{AwadMHSH16,YoungNQ15,SwamiRM16,Zuo0ZZG18}. 

While prior designs mainly offer data confidentiality in the disks, NVM, and ADR-based NVM systems~\cite{AwadMHSH16,LiuKRK18,SwamiRM16,YoungNQ15,Zuo0ZZG18,zuo2019supermem,dickens2018strongbox,casey2008impact,khati2017full}, few schemes discuss how to encrypt data in the eADR-based NVM systems when the CPU caches become persistent domain. Although the on-chip caches are security domains~\cite{YanEPRS06,RogersCPS07,AwadMHSH16,LiuKRK18,YeHA18,ZubairA19,AwadYSNZ19}, efficient encryption is still important for the cached data in the eADR-based NVM system for data confidentiality. Specifically, the caches in the eADR domain offer \textit{transient persistence}, which means that these caches themselves are volatile, and the data in caches are guaranteed to be persisted in the \textit{real} persistent device, e.g., NVM, upon crashes. The plaintext data moving from the caches to the unsafe NVM without encryption become vulnerable to attackers. Unfortunately, we observed that the existing CME scheme in the ADR-based memory controller cannot guarantee data confidentiality in eADR-based NVM systems. The encryption module in CME does not work upon crashes due to power down. However, after system crashes, the eADR works and allows the cached data, whether encrypted or not, to be flushed from caches to NVM, thus leading to information leakage. 



In the eADR-based NVM system, upon crashes, the plaintext data in caches are flushed into NVM to achieve data persistence, without the confidentiality guarantee. The eADR-based NVM systems cannot ensure both data persistence and confidentiality at the same time due to the lack of comprehensive analysis upon access patterns and execution models.

The NVM system behavior can be reduced to three types of operations: read, computation, and write. The system \textbf{reads} the source data from NVM, \textbf{computes} the results via the source data, and \textbf{writes} the results into NVM for long-time storage or the next computation. To implement the encryption scheme in eADR-based NVM systems, we model the execution operations in eADR, ignoring software and hardware implementation details. \textbf{\circled{1}All-Operation Model}: eADR supports \textit{data read, write and computation} upon crashes via the backup battery. \textbf{\circled{2}Write-Compute-Order Model}: eADR supports \textit{data write, and computation} upon crashes via the backup battery. \textbf{\circled{3}Write-Only Model}: eADR only supports \textit{data write} upon crashes via the backup battery. We do not consider the Read-Write-Order Model, since reading data is meaningless without processing the read data. It is worth noting that currently All-Operation Model and Write-Compute-Order Model are \textit{ideal} in theory, and only the Write-Only Model can be supported by current available eADR \textit{in practice}~\cite{eADR}. Existing work~\cite{Horus} has leveraged the ideal models, and we believe the eADR can support these ideal models in the near future.

To efficiently address the dilemma between data persistence and confidentiality in the eADR-based NVM systems, we design different schemes under different eADR execution models. Under the All-Operation Model, existing CME ensures data confidentiality. Under the Write-Compute-Order Model, we introduce the \textbf{B}attery-\textbf{B}acked \textbf{E}ncryption (BBE) scheme via the observation that the backup battery in eADR can support the AES encryption engine and XOR gates upon crashes. We leverage the on-chip incremental counter and outside-the-memory-space address to generate the OTP that is used to XOR with the cached data for encryption upon crashes. Under the Write-Only Model, we further propose the \textbf{Sep}arate \textbf{Encr}yption (Sepencr) scheme that generates the OTPs for all cached data in advance and stores the OTPs on chip. The pre-generated OTPs are XORed with the cached data to complete the encryption in case the system crashes. These encrypted data are finally flushed into NVM.


To evaluate the performance of our proposed designs, we use Gem5~\cite{2020gem5} to implement BBE and Sepencr and run 5 persistent workloads, which are widely used in existing PM designs~\cite{CoburnCAGGJS11,RenZKCWM15,KolliRDSPLCW16,KolliGSDCNW17,LiuKRK18,zuo2019supermem,huang2021star}. Compared with the intuitive in-cache encryption scheme in eADR-based NVM systems, our Sepencr/BBE significantly reduces performance overheads from 403\% to 31\%/4\% under different eADR execution models, respectively.

In summary, this paper makes the following contributions:
\begin{itemize}[leftmargin=*]
		\setlength{\itemsep}{0pt}
		\setlength{\parsep}{0pt}
		\setlength{\parskip}{0pt}
	\item \textbf{Comprehensive modeling of eADR execution from ideal to practice.} We analyze three different eADR execution models from ideal to practice. The All-Operation and Write-Compute-Order Models are ideal, and the Write-Only Model is practical and supported by the current eADR mechanism, which work together to construct comprehensive models for eADR-based systems.
	\item \textbf{The analysis of dilemma between data persistence and confidentiality in eADR-based NVM systems.} We observed that existing eADR-based NVM systems cannot ensure both data persistence and confidentiality at the same time since the CME does not work after crashes while the cached data are still flushed into NVM via eADR, thus causing information leakage.
	\item \textbf{The encryption schemes under different eADR execution models.} To encrypt data in eADR-based NVM systems with low overheads, we leverage the battery of eADR to support the AES engine upon crashes under the ideal eADR execution models. We also pre-generate the OTPs to encrypt the cached data under the practical eADR execution model.
	\item \textbf{Extensive experiments and analysis.} We implement and evaluate our proposed designs. The experimental results show that our BBE and Sepencr significantly reduce performance overheads compared with an intuitive in-cache encryption scheme. We also discuss the applicability issues in eADR-based NVM systems.
\end{itemize}


%% file: background.tex
\section{Backgrounds}
\label{section2}

\subsection{Threat Models}
\label{Threat-Models}

Like the threat models in existing designs~\cite{YanEPRS06,RogersCPS07,AwadMHSH16,LiuKRK18,YeHA18,ZubairA19,AwadYSNZ19}, we assume that only the on-chip domain, including the processor, caches and memory controller (MC), in the computer system is safe. In our threat model, attackers can reveal the data by snooping the memory bus and physically stealing the non-volatile DIMM. The data integrity attacks~\cite{sung1997shared} are beyond the scope of our paper like existing works~\cite{AwadMHSH16,LiuKRK18,SwamiRM16,YoungNQ15,Zuo0ZZG18,zuo2019supermem}, which can be defended via Merkle tree based designs~\cite{GassendSCDD03,RogersCPS07,CostanD16,TaassoriSB18}.

\begin{figure}[b]
	\centering
	\includegraphics[width=0.45\textwidth]{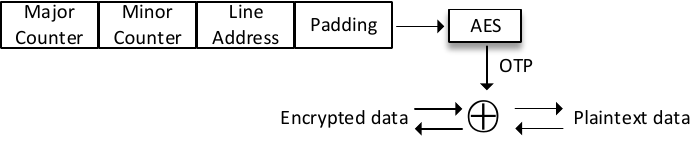}
	\vspace{-0.1cm}
	\caption{The Counter Mode Encryption scheme.}
	\label{CME}
\end{figure}

\vspace{-0.1cm}
\subsection{Counter Mode Encryption}
\label{cme}
To ensure data confidentiality, counter mode encryption (CME) has been widely used in existing NVM-based secure systems~\cite{AwadMHSH16,YoungNQ15,SwamiRM16,Zuo0ZZG18}. The CME is executed in the memory controller and becomes transparent to applications. As shown in Fig.~\ref{CME}, the AES engine encrypts the counter, data memory address and padding via the secret key to generate the one-time padding (OTP). When writing data into NVM, the plaintext data XOR the OTPs to generate the encrypted data. When reading data from NVM, the encrypted data XOR the OTPs to generate the plaintext data. 

In order to guarantee security, the OTPs are one-time and cannot be reused. The inputs for generating OTPs include data memory addresses and counters. For different data lines, since the memory addresses of data are different, the OTPs are different. For the same data line, the corresponding minor counter in Fig.~\ref{CME} increases by 1 when persisting the data line into NVM. Therefore, the OTPs for the same data line on different memory writes are different. The counter block is formed via one 64-bit major counter and 64 7-bit minor counters~\cite{zuo2019supermem,Zuo0ZZG18}. When one minor counter in the counter block overflows, the major counter increases by 1. All 64 minor counters then are reset to 0 and all corresponding 64 data blocks are read to be re-encrypted via the new counters. Compared with direct AES encryption using the unchanged secret key, CME is more secure since the OTPs for encryption are \textit{one-time}. Moreover, since counter blocks are cached in the memory controller, systems generate the OTPs in parallel with reading the encrypted data from NVM. The decryption latency in CME is hidden by the latency of reading data.

\begin{figure}[t]
	\centering
	\includegraphics[width=0.48\textwidth]{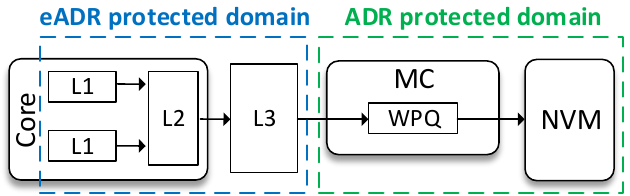}
	\caption{eADR and ADR provide the protection for power failures.}
	\label{eADR}
	\vspace{-0.1cm}
\end{figure}

\vspace{-0.1cm}
\subsection{Integrity Tree}
Data integrity is important to NVM, which means that the data in NVM cannot be modified by attackers. In general, integrity trees are used to fast detect data integrity~\cite{ZubairA19,huang2021star,AwadYSNZ19}. Specifically, a Merkle Tree (MT)~\cite{GassendSCDD03} is a typical integrity tree that is constructed by iteratively hashing the protected user data. A hash value stored on chip is finally generated as the root of MT. In the secure NVM systems with counter mode encryption (CME), a Bonsai Merkle Tree (BMT)~\cite{RogersCPS07} is proposed by coalescing with CME. Unlike MT that iteratively hashes the user data to construct the tree, BMT iteratively hashes the counter blocks of user data in CME to generate the tree root. The user data are protected by the HMACs that are constructed by hashing the user data and their counter blocks. Since the number of counter blocks is less than that of user data, the BMT is smaller than MT with lower storage overheads. MT and BMT are non-parallelizable integrity trees~\cite{ZubairA19}, i.e., in MT/BMT, the parent node is computed after the child node has been computed since the child node is the input of the parent node. To improve the system performance, the parallelizable integrity trees, i.e., SGX-style integrity trees (SIT)~\cite{ZubairA19,alwadi2019phoenix,huang2021star}, are proposed. In SIT, one tree node consists of 8 counters and 1 HMAC. The HMAC in one SIT node is computed by hashing all counters in this node and one counter in the parent node. Therefore, when counters in different nodes have been updated, the HMACs in SIT nodes can be computed in parallel.

\vspace{-0.1cm}
\subsection{eADR Mechanism}

\begin{figure}[t]
	\centering
	\includegraphics[width=0.45\textwidth]{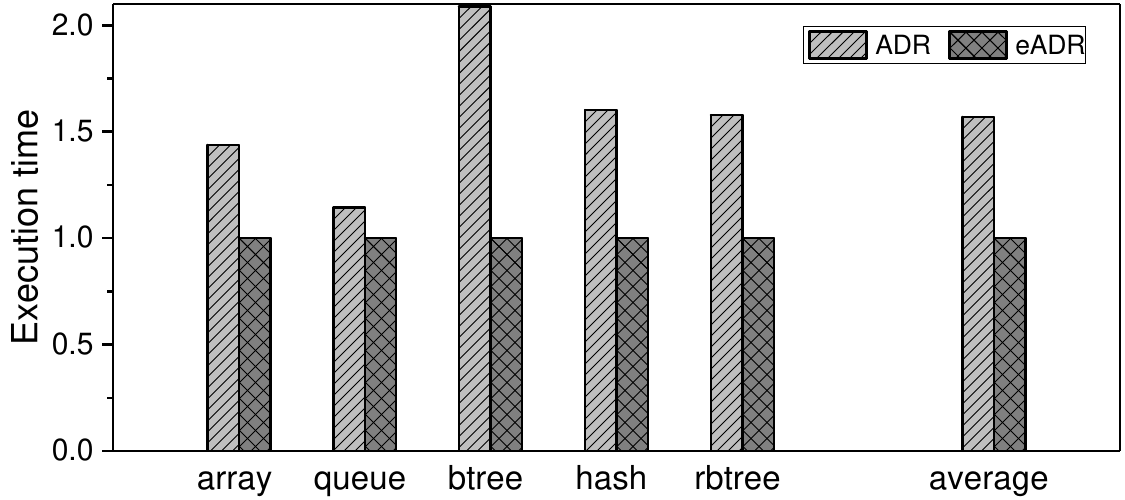}
	\vspace{-0.1cm}
	\caption{The performance of ADR and eADR systems (normalized to eADR).}
	\label{eADR-performance}
	\vspace{-0.2cm}
\end{figure}

\begin{figure*}[t]
	\centering
	\includegraphics[width=0.8\textwidth]{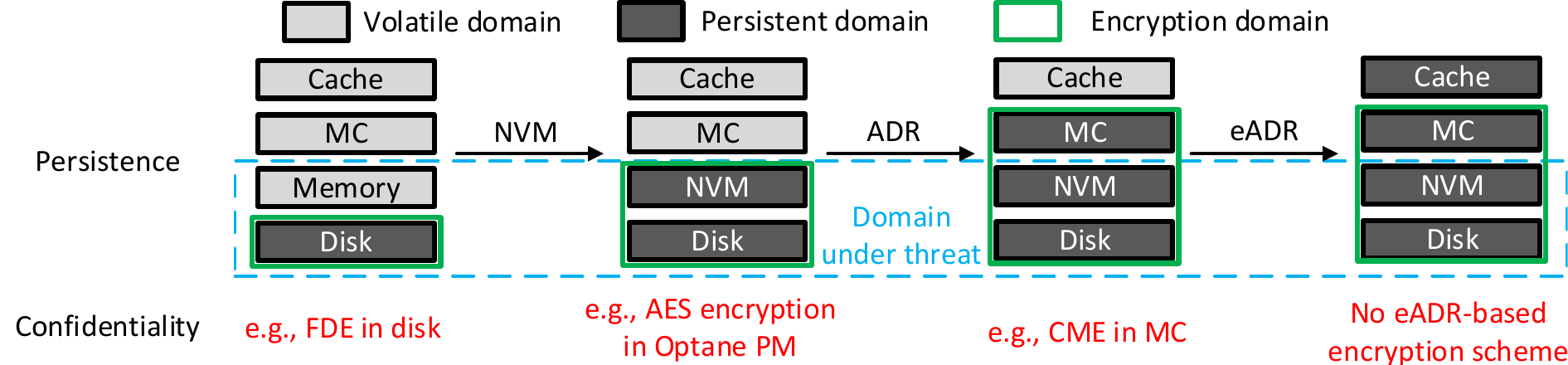}
	\vspace{-0.3cm}
	\caption{The evolution of persistent domain and encryption schemes in different memory architectures.}
	\label{persistence-boundary}
	\vspace{-0.3cm}
\end{figure*}

Intel Asynchronous DRAM Refresh (ADR) technique~\cite{ADR} is able to flush data from the write pending queue (WPQ) in the memory controller into NVM upon system crashes. The WPQ thus becomes a persistent domain. Intel recently proposes extended ADR (eADR)~\cite{eADR} to further extend the persistent domains and improve the system performance. As shown in Fig.~\ref{eADR}, in the eADR environment, the on-chip caches are included in the power-failure protection domain. If system crashes occur, the data in caches can be written into NVM via the supports of eADR~\cite{eADR}. By integrating ADR with eADR, the on-chip domain, including the WPQ and caches, is regarded as the persistent domain. By using the eADR, the flush instructions, such as CLWB, CLFLUSHOPT, or CLFLUSH, become not very necessary, which can be simplified in the context of the non-volatile memory programming~\cite{alshboul2021bbb,yihtmfs,dang2022nvalloc}. Simplified instruction semantics are efficient and helpful to improve system performance. 


As shown in Fig.~\ref{eADR-performance}, we evaluate different workloads in ADR and eADR systems. The system configurations and workloads are described in~\secref{evaluation}. Note that we remove the CLWB, CLFLUSHOPT, and CLFLUSH instructions from source codes to build the eADR-based workloads. This approach to building eADR-based workloads is widely used and well-recognized in existing eADR-based NVM schemes~\cite{alshboul2021bbb,yihtmfs,dang2022nvalloc}. Since the data do not need to be flushed into NVM, eADR significantly improves the system performance. The quantitative performance improvements depend on the number of writes in the workloads. On average, the eADR system improves 57\% system running time compared with the ADR system.


\begin{table}[t]
	\footnotesize
	\vspace{-0.2cm}
	\caption{\label{table:models}The different eADR execution models.}
	\vspace{-0.3cm}
	\begin{center}
		\begin{tabular}{|c|c|c|c|}
			\hline
			\diagbox{Model}{Operation} & Read & Compute & Write        \\
			\hline
			All-Operation Model & $\checkmark$ & $\checkmark$  & $\checkmark$           \\
			\hline
			Write-Compute-Order Model &  & $\checkmark$ & $\checkmark$        \\
			\hline
			Write-only Model &  &  & $\checkmark$            \\
			\hline
			\hline
			\textcolor{red}{Current eADR} &  &  & $\checkmark$				\\
			\hline
		\end{tabular}
	\end{center}
	\vspace{-0.5cm}
	
\end{table}

The battery of eADR is expected to be used to support some operations, including but not limited to writing data upon crashes~\cite{alshboul2021bbb,Horus}. We thus propose three different eADR execution models in this paper from ideal to practice as shown in Table~\ref{table:models}: \textbf{\circled{1}All-Operation Model}: eADR supports \textit{data write, read and computation} upon crashes via the backup battery. \textbf{\circled{2}Write-Compute-Order Model}: eADR supports \textit{data write, and computation} upon crashes via the backup battery. \textbf{\circled{3}Write-Only Model}: eADR only supports \textit{data write} upon crashes via the backup battery. The Write-Only Model is the only model supported by the current version of eADR. Although All-Operation Model and Write-Compute-Order Model are ideal, we believe that they can be supported by eADR in the near future.

\subsection{Counter Crash Consistency}
Systems need to use correct counters to decrypt the encrypted data. While the encrypted data are persisted in NVM, the updated counters are cached in the memory controller for performance improvements. After system crashes, the cached counters in the memory controller are discarded. The stale counters in NVM become inconsistent with the user data and cannot decrypt the data after system recovery. 

Existing designs propose different approaches to ensuring the counter crash consistency~\cite{YeHA18,LiuKRK18,zuo2019supermem}. Supermem~\cite{zuo2019supermem} leverages the write-through counter cache to ensure counter crash consistency. When counters in the counter cache are updated, the modified counters are directly persisted into NVM. Osiris~\cite{YeHA18} does not force to persist the counter blocks with user data. The stale counter blocks can be restored by increasing the counter values. In eADR-based NVM systems, the on-chip domain, including the counter buffer, is the persistent domain via the support of eADR. The counters in the counter buffer can be persisted into NVM upon crashes. Therefore, it is simple to achieve counter crash consistency in eADR-based NVM systems.

\vspace{-0.1cm}
\section{The dilemma between persistence and confidentiality}
\label{section3}

\subsection{The Changes of Persistence Boundary}

In order to store data, hierarchical architectures generally consist of cache, memory, and storage, which have dramatically changed over the decades. As shown in Fig.~\ref{persistence-boundary}, the bottom-up hierarchy contains more and more persistence levels, which meantime introduce associated encryption schemes to protect data in the persistent domain. Specifically, the disk-based persistent domain encrypts data via full disk encryption (FDE)~\cite{dickens2018strongbox,casey2008impact,khati2017full}. Moreover, the non-volatile memory (NVM) is available in the market~\cite{PMEncryption}. When using NVM, the memory is the persistent domain, i.e., the persistence boundary moves up. To protect the data in the NVM, a standard 256-AES hardware encryption~\cite{PMEncryption} is used.

To efficiently offer system consistency, Intel proposed ADR~\cite{ADR}, which is often used with NVM, to flush data from the WPQ in the memory controller into NVM upon system crashes. The memory controller thus becomes the persistent domain in the ADR-based NVM system. To protect the data transiting in the memory bus, which exists in the threatened domain in the context of our threat model (\secref{Threat-Models}), the ADR-based NVM system leverages the CME~\cite{AwadMHSH16,YoungNQ15,SwamiRM16,Zuo0ZZG18} in the memory controller to encrypt the data. As shown in Fig.~\ref{persistence-boundary}, when the persistent domain expands from the disk to the memory controller, the encryption domain also expands to match the persistent domain for data confidentiality.

Recently, the eADR~\cite{eADR} technique is further proposed to extend the persistent domain, in which the data caches become the persistent domain as shown in Fig.~\ref{persistence-boundary}. However, the encryption domain is not extended, causing a mismatch between the persistent domain and the encryption domain. In the eADR-based NVM system, although the persistence boundary moves up compared with the ADR-based NVM system, there are no efficient encryption schemes to protect data.

\vspace{-0.1cm}
\subsection{Data Confidentiality Issues}
In our threat model, the on-chip caches are in the safe domain, and the cached data cannot be attacked by attackers. However, in the eADR-based NVM system, the cached data may be attacked when they are flushed into NVM. Specifically, from the persistence view, the eADR-based caches offer \textit{transient persistence} and require out-of-place flushing to guarantee persistence. The data in cache cannot be persisted in-place upon crashes but need to be flushed into the out-of-place persistent devices, e.g., NVM. To store the data, the system flushes the cached data into NVM upon crashes via the support of eADR. From the security view, the cached data are flushed from the safe domain, i.e., the cache, to the unsafe domain, i.e., the memory bus and NVM (\secref{Threat-Models}). Without data encryption, the plaintext data under the unsafe domain cause information leakage.

\begin{figure}[t]
	\centering
	\includegraphics[width=0.40\textwidth]{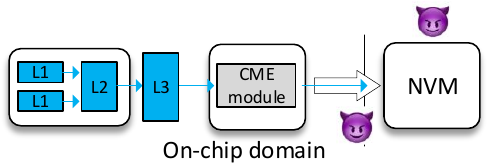}
	\caption{The CME module does not work when system crashes occur. However, the system still flushes the plaintext data from caches into NVM after crashes.}
	\label{delimma}
	\vspace{-0.3cm}
\end{figure}

We argue that existing encryption schemes cannot address the data confidentiality issues in eADR-based NVM systems under the Write-Compute-Order and Write-Only eADR models. As shown in Fig.~\ref{persistence-boundary}, the FDE and standard 256-AES hardware encryption in disk and PM cannot protect data in the memory bus. The CME in the memory controller protects the data passing through the memory bus, but CME is inefficient in eADR-based NVM systems upon crashes. As shown in Fig.~\ref{delimma}, the CME in the memory controller does not work due to counter cache miss and the disability of the encryption engine upon crashes with power failure. But the cached plaintext data are still flushed into NVM for persistence via the support of eADR. The plaintext data may be attacked in the memory bus, i.e., causing data confidentiality issues. In Fig.~\ref{delimma}, the eADR-based NVM system supports data persistence in the CPU caches and however, fails to guarantee data confidentiality.


We observed the dilemma between data confidentiality and persistence in the eADR-based NVM system. Once system crashes, if flushing the cached plaintext data into NVM via eADR, the data confidentiality cannot be guaranteed due to no data encryption. On the other hand, if we discard the cached data upon crashes for data confidentiality, the eADR fails to support data persistence since the data are not flushed into NVM, which invalidates the eADR.

\subsection{The Requirements of Secure eADR-based NVM Systems}
\label{requirements}
In this section, we describe the requirements of a secure eADR-based NVM system that contains persistent off-chip memory and on-chip caches via the support of eADR.

\textit{Requirement 1:} \textbf{Data Confidentiality Requirement.} All user data on the off-chip domain need to be encrypted. As shown in~\secref{Threat-Models}, in our threat model, only the on-chip domains are secure. The off-chip domains, including the memory bus and NVM, are vulnerable to attackers. Since the user data contains sensitive information, all user data in the memory bus and NVM need to be encrypted. Note that the secure metadata do not need to be encrypted, e.g., the counter blocks in the CME and tree nodes in the integrity tree~\cite{TaassoriSB18,huang2021star}.

\textit{Requirement 2:} \textbf{Data Persistence Requirement.} All data in caches need to be persisted into NVM upon system crashes. Since the data in caches can be flushed into NVM via the support of eADR, eADR-based NVM systems remove the CLFLUSH, CLFLUSHOPT, and CLWB instructions and do not need to consider security metadata crash inconsistency. In order to efficiently leverage the persistent caches, many designs remove these flush instructions in the workloads~\cite{alshboul2021bbb,yihtmfs,dang2022nvalloc}. If the data in caches are not persisted upon crashes, the eADR-based workloads are incorrectly running.



%% file: design.tex
\vspace{-0.1cm}
\section{System Designs and Implementations}
\label{design}

%

To ensure data confidentiality, in this section, we first demonstrate an intuitive idea to directly encrypt the data in cache, which however incurs high performance overheads. In order to decrease the overheads, we further present the adaptive encryption schemes that flexibly meet the requirements (\secref{requirements}) of a secure eADR-based NVM system under different eADR execution models. 



\begin{figure}[t]
	\centering
	\includegraphics[width=0.45\textwidth]{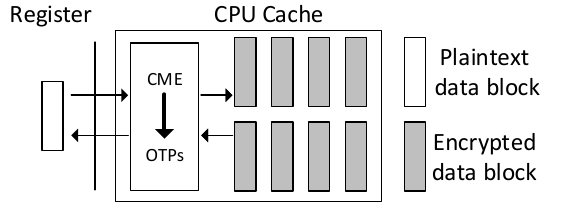}
	\caption{The encryption engine exists in cache to encrypt the cached data.}
	\label{Encryption-In-Cache}
	\vspace{-0.3cm}
\end{figure}

\vspace{-0.2cm}
\subsection{Move the Encryption Up}
\label{base}

To address the dilemma between persistence and confidentiality in the eADR-based NVM system, an intuitive idea is to move the encryption engine to cache to directly encrypt data as shown in Fig.~\ref{Encryption-In-Cache}. The OTPs are generated in cache, and all data in cache are encrypted by XORing the corresponding OTPs. After system crashes, the encrypted data in cache are directly flushed into NVM. However, directly encrypting the data in cache is inefficient for protecting data confidentiality. When a processor reads data from cache, the data need to be decrypted, i.e., generating the OTPs and XORing the data with the OTPs. The data decryption/encryption in the latency-sensitive caches exists on the critical path of reading/writing data, and significantly decreases the system performance as shown in~\secref{evaluation}. Moreover, the cache is designed to store data, which is typically constructed via SRAM~\cite{aly2007low}. Moving the AES engine into cache will increase the complexity of the manufacturing process. 

\vspace{-0.1cm}
\subsection{Data Confidentiality under All-Operation Model}
All-Operation Model is the most relaxed eADR execution model that allows the system to continue reading, writing and processing data upon crashes, like the Uninterruptible Power Supply (UPS) system~\cite{aamir2016uninterruptible}. Under this model, ensuring data confidentiality is simple since we use existing CME to encrypt cached data by reading the security metadata from NVM, and generating and XORing the OTPs in the memory controller upon crashes.

\vspace{-0.1cm}
\subsection{BBE under Write-Compute-Order Model}
\label{sectionBBE}
Write-Compute-Order Model, which allows fewer operations, is stricter than the All-Operation Model. Under Write-Compute-Order Model, the system can write data and continue to run the encryption module upon crashes. We propose a \textbf{B}attery-\textbf{B}acked \textbf{E}ncryption (BBE) to encrypt the data upon crashes to meet both Data Confidentiality and Data Persistence Requirements (\secref{requirements}) under the Write-Compute-Order eADR execution model. We observed two challenges for encrypting cached data after crashes: \circled{1}The loss of counters. The counters are partially stored in the metadata cache for high performance. However, due to the limited size of the metadata cache, counter misses may occur and further prevent the OTP generation upon crashes. Although the system can read counters from NVM to metadata cache during running time, the Write-Compute-Order Model does not support read data upon crashes. \circled{2}The inefficiency of the encryption engine. After system crashes, the power is down. Without power, the AES encryption engine cannot generate the OTPs for encryption. To address these challenges, BBE introduces an \textit{increment counter (incr-counter)} register, which provides counters for the encryption engine upon crashes, and leverages the battery of eADR to support the execution of the encryption module.

\begin{figure}[t]
	\centering
	\includegraphics[width=0.47\textwidth]{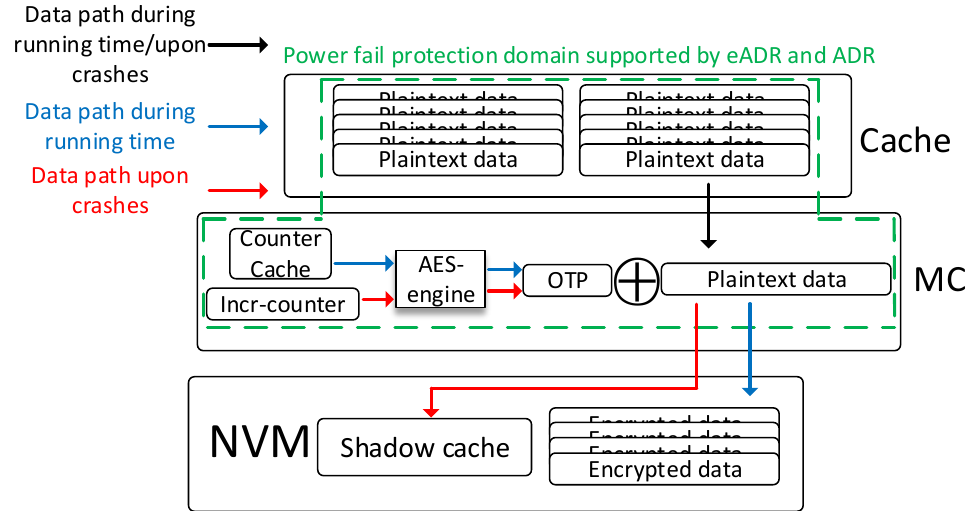}
	\caption{The Battery-Backed Encryption (BBE) scheme.}
	\label{BBE}
	\vspace{-0.5cm}
\end{figure}

\subsubsection{The inputs of CME upon crashes}
The inputs of the AES engine to generate OTPs for memory data are counters, data addresses and paddings. The paddings are padded after counters and data addresses to ensure that the OTP size is the same as the memory line size (64B). The paddings are never changed during running time. We now discuss the counters and data addresses. As shown in Fig.~\ref{BBE}, during running time, the counters are cached in the counter cache, and the system reads the counters from NVM into the counter cache. The incr-counter is stored in the 64b non-volatile register, which is used upon crashes. The initial value of incr-counter is 1. When the incr-counter value is used to generate the OTP, this value increases to ensure that the counters of OTPs are different. Specifically, we assume there is a 4MB cache with 65,536 cache lines. Upon crashes, these cache lines are flushed to the memory controller for encryption, and finally written into NVM one by one via the support of eADR. In the memory controller, the initial incr-counter is leveraged to generate the OTP for encrypting the 1st cache line. The incr-counter then increases to generate the OTP for encrypting the 2nd cache line. Finally, the value of incr-counter is 65,536 for encrypting the 65,536th cache line. After system recovery, the initial value of incr-counter is 65,537 to ensure that the OTPs are never reused. Note that the max value of incr-counter in the 64b non-volatile register is $2^{64}$. Only after $2^{48}$ times of system crashes, the incr-counter overflows. Therefore, we argue that the incr-counter never overflows during the system lifetime.

The data addresses are one of the inputs to generate the OTPs. To ensure the OTPs are never reused, we need to use the addresses, which are different from those used during the running time, to generate the OTPs upon crashes. Specifically, for one particular cache line, since the incr-counter increases from 1, the incr-counter for the cache line may be the same as the counter which is used to encrypt the line during running time. If the data address for generating the OTP is unchanged, the OTP of the line may be reused, thus violating the one-time principle (\secref{cme}). To guarantee that the OTPs are never reused, we use the outside-the-memory-space address to generate the OTPs. Assuming the NVM is 16GB, there are many unused memory addresses in the 64-bit address space. For the 16GB NVM, we use 16GB$+$$N$$\times$$64$B as the address of the $N$th cache line to generate the OTP upon crashes. These outside-the-memory-space addresses are never used during system running time. Therefore, we ensure that the OTPs used upon crashes are different from those used during running time.

Generating the OTPs incurs the write latency and decreases the system performance, which is addressed by existing works~\cite{han2021dolos,liu2019janus}. Moreover, since system crashes are rare, the high write latency upon crashes does not affect the entire system performance.


\subsubsection{The encryption engine with backup battery}
After system crashes, the encryption engine without power cannot continue running. To support the generation of OTPs upon crashes, as shown in Fig.~\ref{BBE}, we place the AES engine and the XOR gate in the eADR protection domain. The AES engine includes more than ten rounds of SubBytes, ShiftRows, MixColumns and AddRoundKey~\cite{giraud2004dfa} to encrypt data. Fortunately, we found that the modern AES engine is cost-efficient, e.g., DW-AES engine~\cite{wang2014energy} achieves an energy efficiency of 24 $pJ/bit$ with about 78.121 $um^{2}$ area costs. The XOR gate is one of the basic circuits with the simplest operation~\cite{gate}. Moreover, the silicon area and energy overheads of the XOR gate are very small, i.e., about 16.56 x 12.81 $um^{2}$ with 11 transistors, and 100 $fJ/byte$~\cite{saravanan2015energy}. We thus leverage the backup battery to support the encryption engine to generate the OTPs and XOR gate to XOR the data and OTPs under the eADR Write-Compute-Order Model upon crashes. We analyze the energy consumption of BBE scheme in~\secref{evaluation}. The shadow cache in NVM in Fig.~\ref{BBE} demonstrates the same structure as the on-chip cache, and is used to store the cached data one by one upon crashes, e.g., the $N$th memory line in the shadow cache stores the $N$th cache line after crashes.

\subsubsection{The work flow of BBE}
There are two states in the eADR-based system: system running time and system crashes. As shown in Fig.~\ref{BBE}, during system running time, the cached data is encrypted by CME and flushed into its corresponding address in the NVM. After crashes, the AES engine generates the OTPs via the incr-counter. The system further XORs the OTPs with the cached data to encrypt the data via the XOR gate. The AES engine and XOR gates are supported by the backup battery of eADR upon crashes under the Write-Compute-Order Model. When the cached data is encrypted, the data is flushed into the shadow cache in the NVM, instead of the corresponding address, since the encryption way (i.e., the BBE) of this data is different from other data (i.e., the conventional CME) in NVM.

During system recovery, the system reads the data from the shadow cache to the memory controller and decrypts them. Since the structure of the shadow cache in NVM is the same as the data cache, we can obtain the correct counter by checking the location of the data in the shadow cache and the current incr-counter. All the data in the shadow cache are read, decrypted, and stored in the cache. The system then continues running.

In BBE, in addition to writing data into NVM upon crashes, we leverage the battery of eADR to support the running of the AES engine and XOR gate under the Write-Compute-Order Model of eADR.

\begin{figure}[t]
	\centering
	\includegraphics[width=0.45\textwidth]{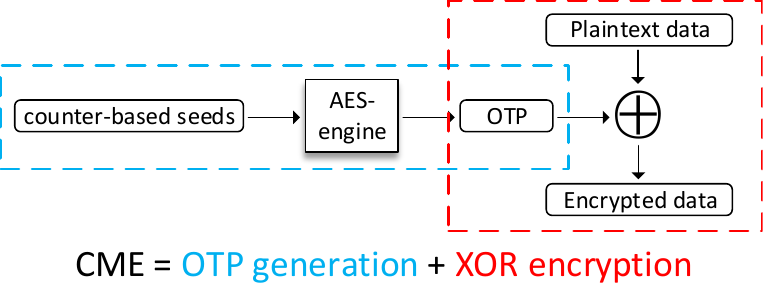}
	\vspace{-0.1cm}
	\caption{The separate structure of counter mode encryption.}
	\label{CME-steps}
	\vspace{-0.3cm}
\end{figure}

\vspace{-0.1cm}
\subsection{Sepencr under Write-Only Model}
\label{sepencr}
Write-Only Model only allows the system to write data into NVM upon crashes, which is the function of the current version of eADR. We propose a \textbf{Sep}arate \textbf{Encr}yption (Sepencr) scheme to ensure data confidentiality in the eADR-based NVM system under the practical Write-Only Model of eADR by generating the OTPs in advance.

\subsubsection{The Separate Structure of CME}
As shown in Fig.~\ref{CME-steps}, CME is partitioned into two stages: OTP generation and XOR encryption. In the OTP generation stage, the counter-based seeds (i.e., counters, addresses and paddings) are encrypted by the secret key via the AES engine to generate the OTPs. In the XOR encryption stage, the OTPs are XORed with plaintext data to perform encryption. In the CME, the OTP generation dominates the latency, and the operation of XORing OTPs with data becomes fast, i.e., less than 1 cycle~\cite{wang2020novel}. The data are finally encrypted in the XOR encryption stage.

From Fig.~\ref{CME-steps}, we observed that CME essentially is a separate encryption structure. The OTP generation and XOR encryption stages can be separately performed. Based on this observation, we propose \textbf{Sep}arate \textbf{Encr}yption (Sepencr) to encrypt data in the eADR-based NVM systems under the practical Write-Only eADR execution model. The idea of Sepencr is to decouple the OTP generation and data encryption (i.e., XORing OTP with data). In our Sepencr, the OTP generation and data encryption are performed in different locations and times.

\subsubsection{Sepencr Overview}
\begin{figure}[t]
	\centering
	\includegraphics[width=0.45\textwidth]{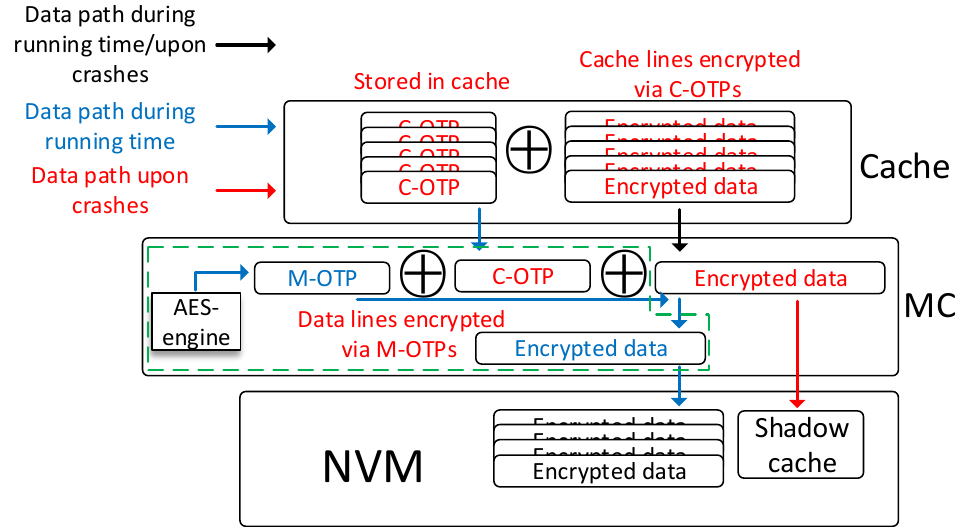}
	\caption{The data are encrypted via C-OTPs/M-OTPs in cache/MC upon crashes/during system running time, respectively. The encryption module in the green dotted line does not work upon crashes.}
	\label{SepencrInCache}
	\vspace{-0.3cm}
\end{figure}

Sepencr generates the OTPs in advance and leverages the pre-generated OTPs to encrypt cached data in case the system crashes. As shown in Fig.~\ref{SepencrInCache}, the pre-generated OTPs are stored in cache (called C-OTPs). Every cache line is encrypted in cache by XORing the cache line with the corresponding C-OTP. In the memory controller, conventional CME is generating the OTPs (called M-OTP) via the memory addresses and counters. During system running time, the encrypted cache line is flushed into the memory controller with the corresponding C-OTP. The encrypted cache line is decrypted by the corresponding C-OTP in the memory controller and re-encrypted by the M-OTP. The re-encrypted data is then written into NVM. Upon crashes, the encrypted cached data are directly flushed into the shadow cache in NVM. The structure of shadow cache in Fig.~\ref{SepencrInCache} is described in~\secref{sectionBBE}.

\subsubsection{Generating C-OTPs }

\textbf{The addresses for generating OTPs.}
When the cached data enters the memory controller, the M-OTP is generated on demand via the memory address of the cached data. However, the memory addresses of cached data are not suitable for generating the C-OTPs due to security issues. Specifically, assuming we leverage the memory addresses of cached data to generate the C-OTPs. When writing a data into cache, the system needs to get the memory address, generate the OTP, move the OTP to cache as the C-OTP, and encrypt the cached data via the C-OTP. If the system crashes occur after the data has been written into cache but before the OTP is generated and XORed with the data, the plaintext data will be flushed into the NVM. Moreover, the memory address used to generate C-OTP may have been used to generate the M-OTP. If the counters of C-OTP and M-OTP are the same (we discuss the counter later), the C-OTP/M-OTP may be reused, which violates the security principle of OTP.

To ensure the cached data are instantly encrypted for confidentiality and avoid address reuses between M-OTPs and C-OTPs, we leverage different types of addresses to generate M-OTPs and C-OTPs like the BBE scheme (\secref{sectionBBE}). For M-OTPs, we still use the memory addresses of the data to-be-encrypted as the inputs of CME. For C-OTPs, we use the outside-the-memory-space addresses as the input of CME. For the $N$th cache line in the CPU cache, we leverage 16GB$+$$N$$\times$$64$B as the address (called C-address) to generate the corresponding C-OTP when the NVM size is 16GB. According to the line's location in the cache, each cache line corresponds to a unique and fixed C-address. 

The C-addresses of all cache lines can be deduced via the locations of the cache lines in the cache. By using the C-addresses, we generate the C-OTPs for all cache lines when the system starts, at which time there are no data in cache. To immediately encrypt the cached data, as shown in Fig.~\ref{SepencrInCache}, we store all C-OTPs in the cache. Consequently, any data to be written into the cache line is XORed with the corresponding C-OTP to be encrypted instantly. When reading data from cache to the CPU register, the cached data are decrypted by XORing the encrypted data with the C-OTPs in the cache. The XOR operation is fast~\cite{wang2020novel} and does not incur high performance overheads as shown in~\secref{evaluation}. Since the cached data in a 64B cache line is XORed with the C-OTP bit by bit, the size of the C-OTP is the same as that of the corresponding cache line. Half of the cache space is hence used to store C-OTPs for encrypting the cached data.

\textbf{When to use C-OTPs.}
However, for the multi-way set-associative cache~\cite{seznec1993case}, multiple 64B data blocks in memory will be cached into the same cache line. Multiple data blocks hence share the same C-address to generate the same C-OTPs, which violates the OTP principle as shown in~\secref{cme}, i.e., different data lines leverage different OTPs. To ensure data security, as shown in Fig.~\ref{SepencrInCache}, we divide the system into two states, i.e., running state and crash state. During system running time, the data are encrypted by M-OTPs. Upon system crashes, the data are encrypted by C-OTPs.

The data blocks in cache are always encrypted by the C-OTPs. However, the data blocks persisted into NVM are encrypted by M-OTPs during running time. Specifically, as shown in Fig.~\ref{SepencrInCache}, when one encrypted cached data block is flushed into NVM, the encrypted data block is XORed with the corresponding C-OTP and M-OTP in the memory controller. The encrypted data block is decrypted by C-OTP to obtain plaintext data block, and then re-encrypted by M-OTP after XORing C-OTP and M-OTP. The data block re-encrypted via M-OTP is further flushed into the NVM. Therefore, during running time, the data in the system are encrypted like existing security NVM systems via CME~\cite{AwadMHSH16,YoungNQ15,SwamiRM16,Zuo0ZZG18}. Upon system crashes, the encryption module (i.e., AES engine and XOR gates) framed by the green dotted line in Fig.~\ref{SepencrInCache} does not work under the Write-Only Model of eADR. The data blocks encrypted via C-OTPs in the cache are directly flushed into NVM via the support of eADR. The C-OTPs are hence only used upon system crashes.


%

\textbf{The counters for generating C-OTPs.}
Sepencr generates the C-OTPs at system startup. Although the cached data are always encrypted via C-OTPs, in fact, the C-OTPs are used only upon system crashes. The new C-OTPs are further generated on system recovery for future use. Consequently, the C-OTPs are \textit{one-time} like the OTPs. During running time, the C-OTPs are not used and do not need to be changed until crashes occur. The counters of C-OTPs increase by 1 only on the system crashes and recovery. Since the initialization values of all counters are 1 and the counters increase at the same time, the values of all counters are the same at any time. We use one 64b non-volatile register on-chip to store the value of counters of C-OTPs, which is also the number of times the system crashes during the system lifetime. Since the 64b counter overflows only after $2^{64}$ times of system crashes, which is impossible during the NVM system lifetime, the 64b register is enough to record the values of counters. Our Sepencr only generates the C-OTPs on system crashes and recovery. The overheads of generating C-OTPs are thus negligible. 

The counters used to generate C-OTPs are different from the incr-counter in BBE. The incr-counter is increased after encrypting one cached data to encrypt the next cached data. In Sepencr, all C-OTPs are generated by the same counter, and the counter is increased only after system crashes and recovery.

%

\subsection{Integrity Trees in eADR-Based Systems}
Integrity trees are used to protect data integrity in NVM systems. When the user data are written into NVM, the integrity tree needs to read the intermediate tree nodes for propagating the modifications to the root by generating the HMACs in the intermediate nodes. Since this propagation process requires reading data upon crashes, it only runs in the ideal All-Operation eADR execution model upon crashes. In the Write-Compute-Order and Write-Only eADR models, when crashes occur, the root is inconsistent with the user data since the modification propagation in the integrity tree interrupts.

Fortunately, existing work SCUE~\cite{huang2021update} proposes a shortcut update scheme to immediately update the integrity tree root from user data by skipping the intermediate tree nodes in SIT, which can be leveraged in the Write-Compute-Order and Write-Only eADR Models. Data integrity is beyond the scope of our confidentiality work, and our design is orthogonal to SCUE to ensure data integrity in eADR-based NVM systems.

%% file: experiment.tex
\section{Performance Evaluation}
\label{evaluation}

\subsection{Evaluation Methodology}
\label{section 5}
To evaluate the performance of BBE and Sepencr, we model the cycle-accurate systems in the Gem5~\cite{2020gem5}. The main parameters are shown in Table~\ref{table:configure}. The counter cache in the memory controller is 512KB for storing counter blocks. Since the integrity verifications~\cite{GassendSCDD03,RogersCPS07,CostanD16,TaassoriSB18} are beyond the scope of our paper, we do not consider the integrity tree nodes in the system. We model the 16GB NVM via PCM technologies. The PCM latency is modeled like well-recognized designs~\cite{zuo2019supermem,xu2015overcoming,huang2021star}. Since the counter mode encryption is not leveraged in DRAM, we model the NVM without DRAM like existing schemes~\cite{AwadYSNZ19,freij2020streamlining,freij2021bonsai,YeHA18,ZubairA19}. We use 5 typical persistent workloads, i.e., array, queue, btree, hash, and rbtree, which are widely used in state-of-the-art NVM schemes~\cite{CoburnCAGGJS11,RenZKCWM15,KolliRDSPLCW16,KolliGSDCNW17,LiuKRK18,zuo2019supermem,huang2021star}, to evaluate the systems. Like existing eADR-based designs~\cite{alshboul2021bbb,yihtmfs,dang2022nvalloc}, we remove the CLFLUSH, CLFLUSHOPT, and CLWB instructions from source codes to build the workloads.

\begin{table}[!tbp]
	\footnotesize
	\vspace{-0.2cm}
	\caption{\label{table:configure}The configurations of the evaluated NVM system.}
	\vspace{-0.3cm}
	\begin{center}
		\begin{tabular}{|c|c|}
			\hline
			\multicolumn{2}{|c|}{\textbf{Processor}} \\
			\hline
			CPU & 8 cores, X86-64 processor, 2 GHz        \\
			\hline
			Private L1i/d cache & 128KB, 2-way, LRU, 64B Block              \\
			\hline
			Shared L2 cache & 1MB, 8-way, LRU, 64B Block        \\
			\hline
			Shared L3 cache & 2MB, 8-way, LRU, 64B Block            \\
			\hline
			\hline
			\multicolumn{2}{|c|}{\textbf{DDR-based PCM Main Memory}} \\
			\hline
			Capacity &  16GB     \\
			\hline
			PCM latency model &   \tabincell{c}{tRCD/tCL/tCWD/tFAW/tWTR/tWR \\=48/15/13/50/7.5/300 ns}     \\
			\hline
			Write queue & \tabincell{c}{64 entries}   \\
			\hline
			\hline
			\multicolumn{2}{|c|}{\textbf{Secure Parameters}} \\
			\hline
			Counter cache & \tabincell{c}{512KB, in MC}  \\
			\hline
			C-OTPs & \tabincell{c}{2MB in caches (Sepencr)} \\ 
			\hline
		\end{tabular}
	\end{center}
	\vspace{-0.5cm}
	
\end{table}

To comprehensively examine the performance of our proposed designs, we evaluate and compare the following schemes.
\vspace{-0.1cm}
\begin{itemize}[leftmargin=*]
	\setlength{\itemsep}{0pt}
	\setlength{\parsep}{0pt}
	\setlength{\parskip}{0pt}
	\item Unsecure eADR-based NVM system as Baseline. The Baseline system contains the eADR mechanism without any data encryption, and hence achieves the optimal performance.
	\item The eADR system with CME in cache (eADR-CME). eADR-CME moves the AES engine from the memory controller to the cache (\secref{base}) to encrypt data in the cache. The encryption and decryption operations exist on the critical path of writing/reading data into/from caches. eADR-CME can work on all eADR execution models to ensure data confidentiality.
	\item Our proposed BBE (BBE). BBE leverages the battery of eADR to support the AES engine and XOR gates upon crashes to encrypt the cached data (\secref{sectionBBE}). BBE works under the Write-Compute-Order Model.
	\item Our proposed Sepencr (Sepencr). Sepencr pre-generates the C-OTPs, and stores the C-OTPs in the cache (\secref{sepencr}). Upon crashes, the data encrypted via C-OTPs in the cache are flushed into NVM. The Sepencr works under the Write-Only Model to protect data in eADR-based systems.


\end{itemize}


\begin{figure}[t] \centering
	\subfigure[64B transaction size]{	
		\begin{minipage}[b]{0.45\textwidth}
			
			\includegraphics[width=1\textwidth]{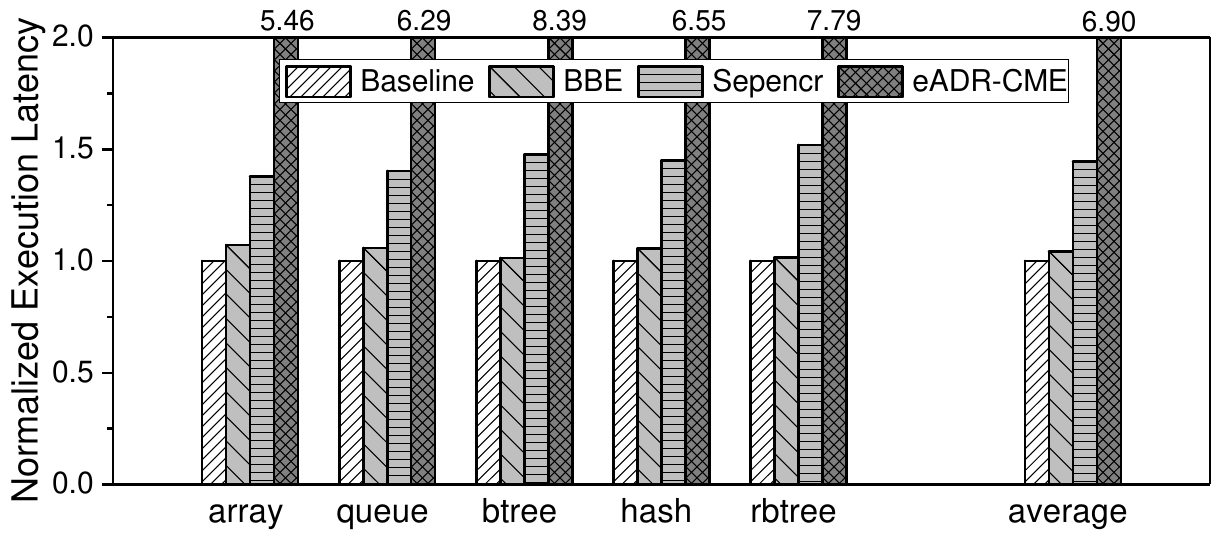}
		\end{minipage}	
	}
	\vspace{-0.3cm}
	
	\subfigure[256B transaction size]{
		\begin{minipage}[b]{0.45\textwidth}
			
			\includegraphics[width=1\textwidth]{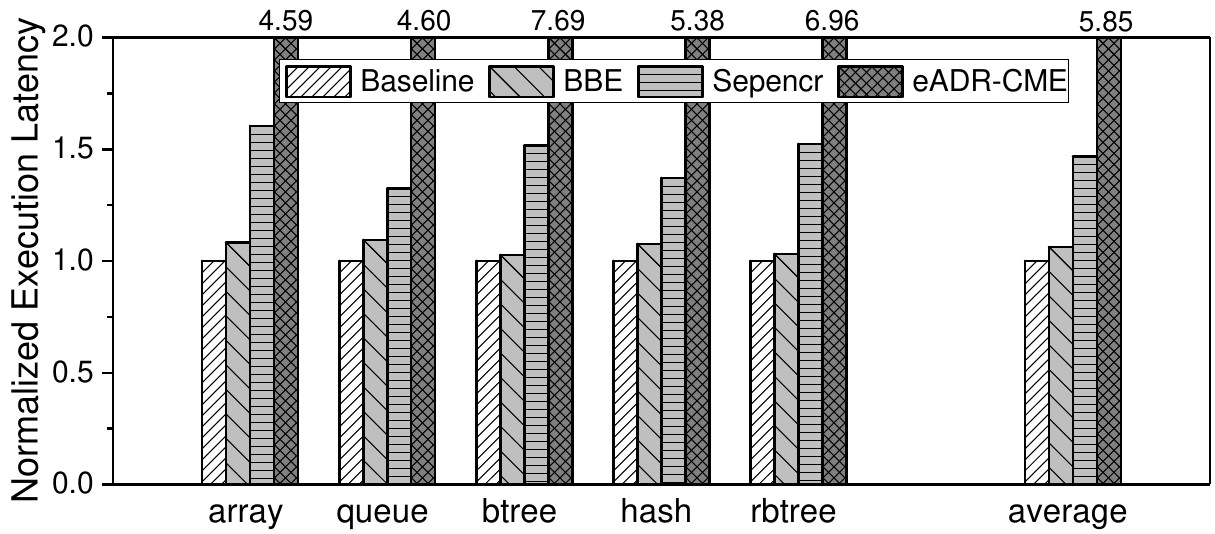}
		\end{minipage}	
	}
\vspace{-0.3cm}
	
	\subfigure[1024B transaction size]{
		\begin{minipage}[b]{0.45\textwidth}
			
			\includegraphics[width=1\textwidth]{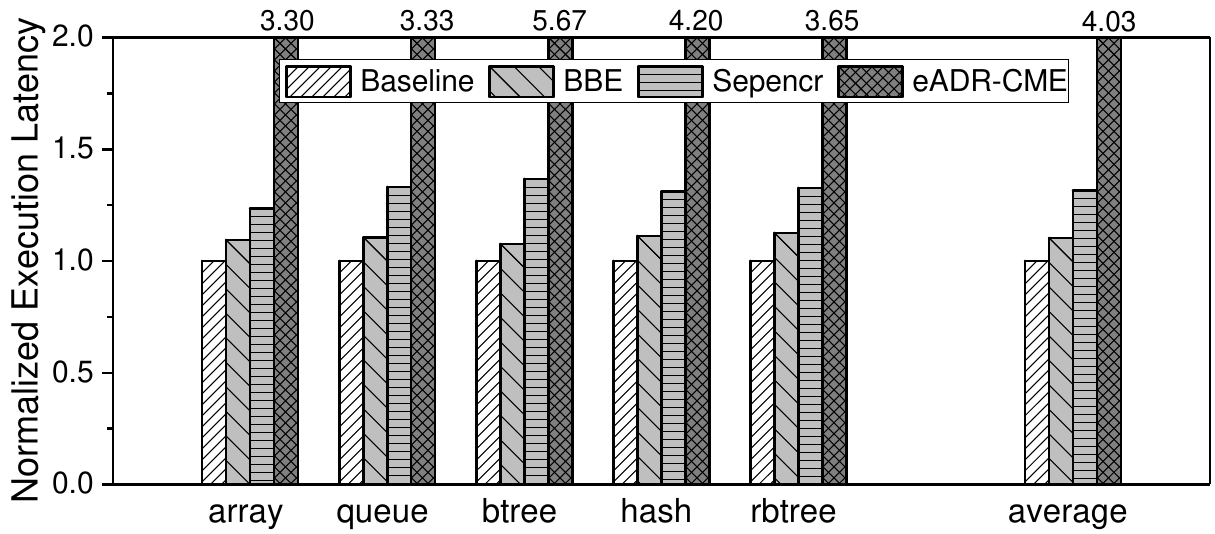}
		\end{minipage}	
	}
	\vspace{-0.1cm}
	\caption{The execution latencies of running workloads in different transaction sizes (normalized to Baseline).}
	\label{single_processor}
	\vspace{-0.4cm}
\end{figure}

\subsection{Result Analysis}
\textbf{The Single-core Performance.}
We run the workloads in different transaction sizes in the schemes, e.g., from 64B to 1024B. As shown in Fig.~\ref{single_processor}, the overheads of eADR-CME are very high. On average, the execution latency of eADR-CME is 4.03x -- 6.90x than that of the Baseline scheme. In eADR-CME, when writing data from the processor into the cache, the AES engine generates the OTPs to encrypt the plaintext data. When reading data from the cache to the processor, the AES engine generates the OTPs to decrypt the encrypted data. The encryption/decryption significantly decreases the performance of eADR-CME. Unlike eADR-CME, Sepencr prepares the OTPs at system startup, i.e., the C-OTPs. Sepencr stores C-OTPs in cache and leverages C-OTPs to encrypt cached data in case the system crashes. Since half of the cache in Sepencr is used to store C-OTPs, the cache contention of user data in Sepencr increases the performance overheads. However, the increased transaction sizes also result in the severe cache contention of the Baseline. The performance overheads of Sepencr in some workloads (e.g. the \textit{array} workload), which are normalized to Baseline, decrease when the transaction size becomes 1024B, i.e., the size of 16 cache lines. On the other hand, the XOR operations in latency-sensitive caches in Sepencr decrease the system performance. On average, the execution latency of Sepencr is 1.31x -- 1.46x than that of the Baseline scheme. In BBE, the data are encrypted/decrypted via the traditional CME method during running time. Upon crashes, BBE encrypts the user data with the support of eADR energy via new counters and addresses. Since system crashes are rare, the encryption latencies of BBE upon crashes do not affect system performance. Moreover, since the decryption latency in CME during running time is masked by that of reading data from NVM~\cite{AwadMHSH16,YoungNQ15,SwamiRM16,Zuo0ZZG18}, the execution latency of BBE is small, i.e., 1.04x -- 1.10x on average.

\begin{figure}[t] \centering
	\subfigure[2 programs]{	
		\begin{minipage}[b]{0.45\textwidth}
			
			\includegraphics[width=1\textwidth]{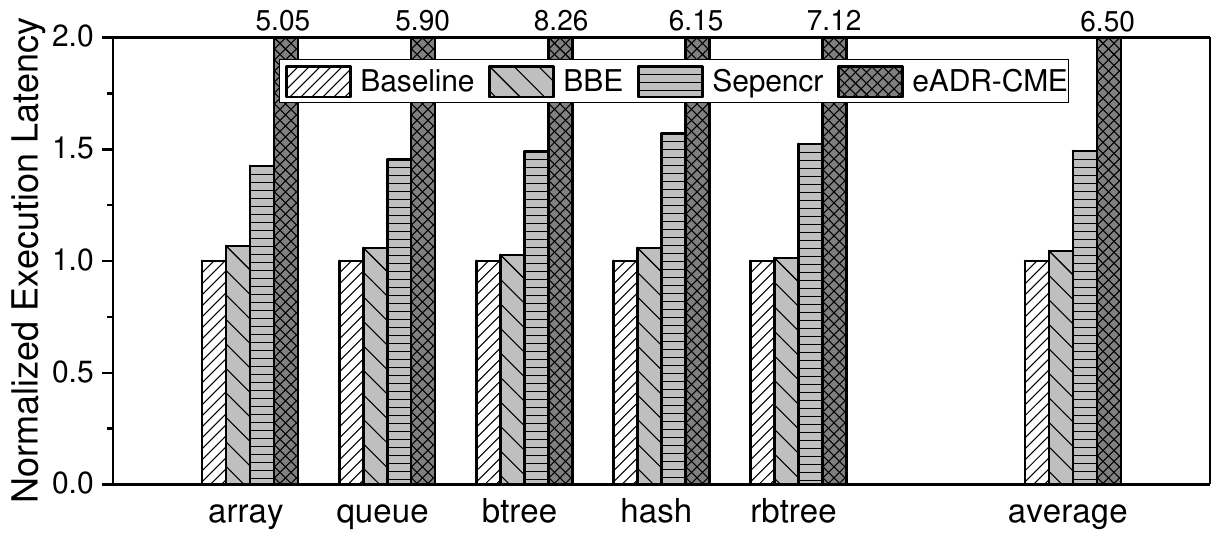}
		\end{minipage}	
	}
	\vspace{-0.3cm}
	
	\subfigure[4 programse]{
		\begin{minipage}[b]{0.45\textwidth}
			
			\includegraphics[width=1\textwidth]{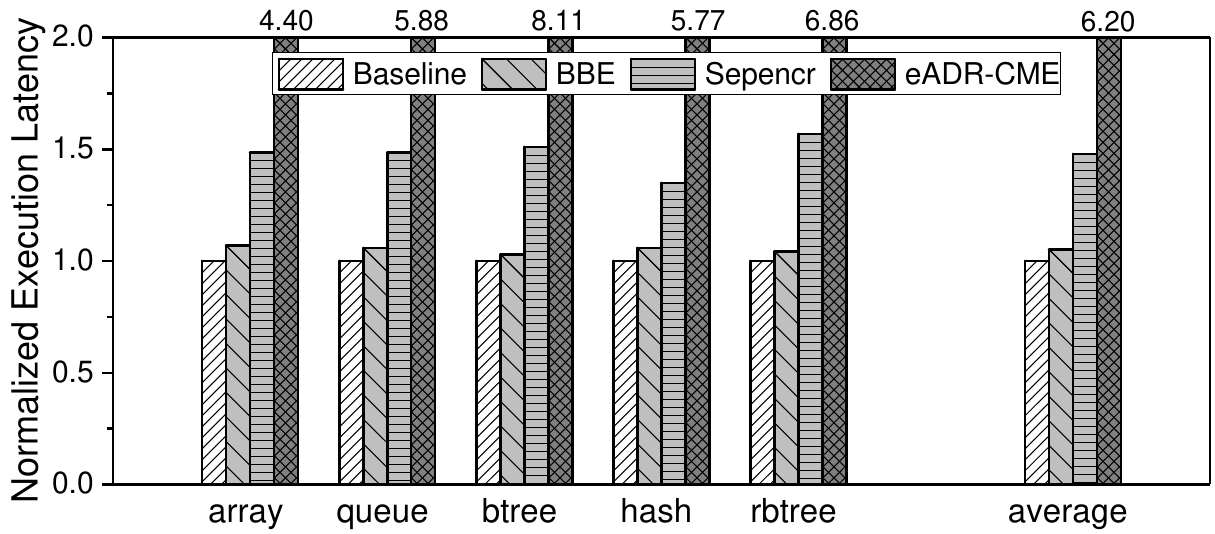}
		\end{minipage}	
	}
\vspace{-0.3cm}

\subfigure[8 programs]{
	\begin{minipage}[b]{0.45\textwidth}
		
		\includegraphics[width=1\textwidth]{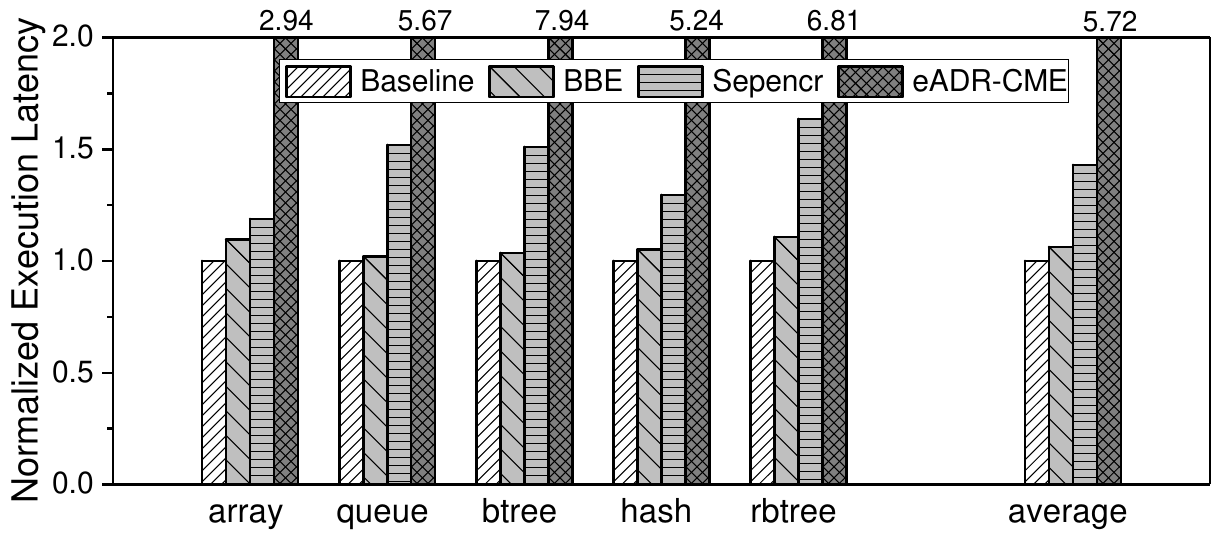}
	\end{minipage}	
}
	\vspace{-0.1cm}
	\caption{The execution latencies in the multi-core systems (normalized to Baseline).}
	\label{multi_processor}
	\vspace{-0.1cm}
\end{figure}

\textbf{The Multi-core Performance.}
To demonstrate the performance impact of our proposed schemes, we run the workloads in a different number of cores (from 1 to 8), where each thread executes the same workload in 64B transaction size on different cores. As shown in Fig.~\ref{multi_processor}, the performance trends of Sepencr are different in the multi-core system for different workloads. For example, from 2 cores to 4 cores, the performance overheads of \textit{array} workload in Sepencr increase since the cache contention in Sepencr is severe when only half of the cache space is used to store user data. In the 8-core system, the performance overheads of Sepencr in the \textit{array} workload significantly decrease. The reason is that the L2 and L3 caches are shared by all cores, which leads to the cache contention when running the \textit{array} workload in Baseline. The performance of Baseline in an 8-core system significantly decreases. On average, the execution latency of Sepencr increases by 1.42x -- 1.47x compared with Baseline.

Unlike Sepencr, the performance of BBE normalized to Baseline is stable, since the cache usage of BBE is similar to that of Baseline. The average execution latency of BBE increases by 1.05x -- 1.07x compared with Baseline.

\subsection{Space and Energy Costs}

The extra space costs in Sepencr over the Baseline scheme are incurred by storing C-OTPs in cache and the encryption module in the memory controller. To execute traditional CME, a 512KB counter cache is used in the memory controller. A 64b register is used to store the values of counters to generate C-OTPs in Sepencr. The size of one C-OTP is the same as that of one data line (64B). To encrypt all cached data, when storing C-OTPs in data caches in Sepencr, half of the cache space is used to store C-OTPs. As shown in Table~\ref{table:configure}, in our configurations, the size of data caches (eight private L1 data, one L2 and one L3 caches) is 4MB (8x128KB + 1MB + 2MB = 4MB). In Sepencr, 2MB space in cache is used to store C-OTPs, and the other 2MB is used to store user data. Moreover, since the atomic operation granularity in the processor is 8b, we place 8 XOR gates in the memory controller in Sepencr. The silicon area costs of 8 XOR gates are 0.001697 $mm^{2}$ with 88 transistors~\cite{saravanan2015energy}. Moreover, the DW-AES engine requires about 78.121 $um^{2}$~\cite{wang2014energy}. For BBE, in addition to these counter cache, AES engine, and XOR gates, BBE also leverages a 64b register to store the incr-counter.

\begin{table}[!t]
	\vspace{-0.3cm}
	\scriptsize
	\caption{\label{EnergyOverhead}The energy costs of Sepencr, BBE and baseline eADR system upon crashes .}
	\vspace{-0.3cm}
	\begin{center}
		\setlength{\tabcolsep}{1mm}{
			\begin{tabular}{|c|c|c|c|c|}
				\hline
				\multirow{2}*{Systems}&\multicolumn{2}{|c|}{Cost of flushing data}&\multirow{2}*{Cost of encryption module}&\multirow{2}*{Overall cost} \\
				\cline{2-3}
				&Caches & Memory controller & &\\
				\hline
				Baseline & 47.7343mJ & / & / & 47.7343mJ \\
				\hline
				Sepencr & 23.8672mJ & 5.8867mJ & / & 29.7539mJ \\
				\hline
				BBE & 47.7343mJ & 5.8867mJ & 0.8087mJ & 54.4297mJ \\
				\hline
		\end{tabular}}
	\end{center}
	\vspace{-0.5cm}
	
\end{table}

We estimate the energy costs of Sepencr and BBE following BBB~\cite{alshboul2021bbb} by using the results from Dhinakaran et al.~\cite{pandiyan2014quantifying}. Since these papers~\cite{pandiyan2014quantifying,alshboul2021bbb} do not include the costs of flushing data from the L3 cache and memory controller into NVM, we assume the cost is the same as that of flushing data from the L2 cache into NVM even though the L3 cache and memory controller are closer to NVM than the L2 cache in the memory hierarchy. The energy costs of flushing data from the L1 data cache/L2 cache/L3 cache/memory controller into NVM are 11.839/11.228/11.228/11.228$nJ/Byte$. Moreover, the energy cost of XORing two bytes to obtain one byte is 800$fJ$~\cite{saravanan2015energy}, and generating OTPs via the DW-AES engine is 192$pJ/Byte$~\cite{wang2014energy}. As shown in Table~\ref{EnergyOverhead}, we estimate the energy cost of flushing data from the caches (L1 data, L2, and L3 caches) and memory controller (counter cache) into NVM upon crashes. We also estimate the energy cost of the encryption module (AES engine and XOR gates) to process the data. We emphasize that the results in Table~\ref{EnergyOverhead} do not accurately demonstrate the energy cost of eADR. However, they show the relative energy costs of different schemes. 

Since only half of the cache stores user data in Sepencr, the energy cost of Sepencr is about half that of BBE, and less than Baseline. Compared with Baseline, BBE needs to drain 512KB counter cache from the memory controller into NVM, and consumes the energy to support the encryption engine and XOR gates upon crashes. BBE incurs 14\% extra energy cost than Baseline due to flushing counter cache. Moreover, it is worth noting that the energy costs of the encryption engine and XOR gates are extremely low compared with the cost of Baseline.

\begin{figure}[!thbp]
	\centering
	\includegraphics[width=0.45\textwidth]{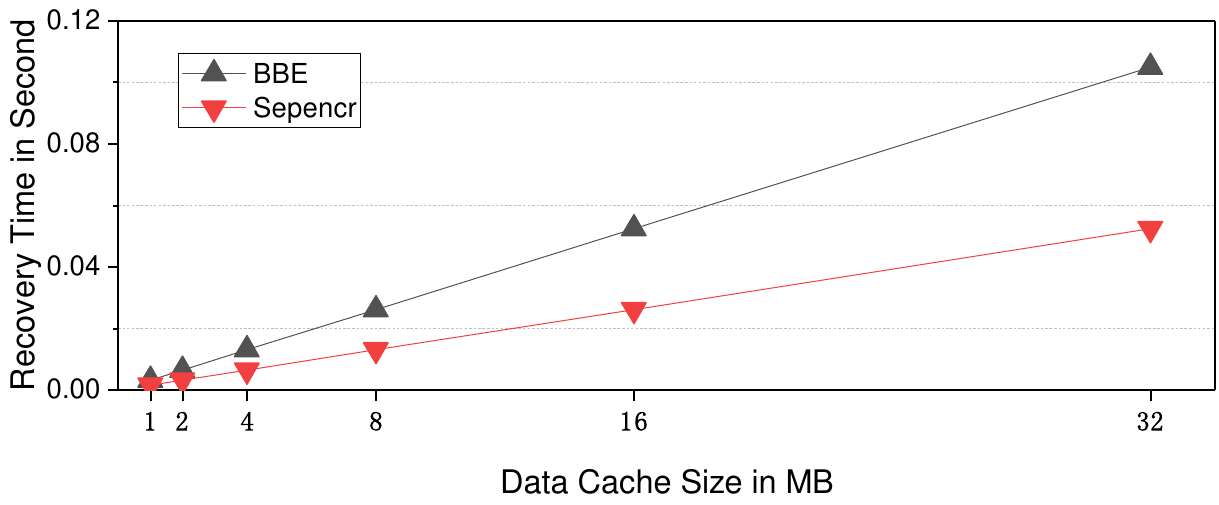}
	\vspace{-0.3cm}
	\caption{The recovery time of BBE and Sepencr for different sizes of data caches.}
	\label{recovery_time}
	\vspace{-0.1cm}
\end{figure}

\vspace{-0.2cm}
\subsection{Recovery Time}
After a reboot from system crashes, since the cached data before crashes are encrypted via BBE/Sepencr while other data in NVM are encrypted via conventional CME, the system needs to first read the cached data and leverage the BBE/Sepencr to generate the OTPs to decrypt the encrypted data. After the data are decrypted, the Sepencr generates the new C-OTPs to encrypt the cached data. Following existing designs~\cite{ZubairA19,huang2021star}, we assume that reading a 64B block from NVM to cache needs about 200ns. Since generating OTPs can be executed in parallel with reading data, the latencies of reading data from NVM dominate the recovery time.

The recovery times of different schemes are shown in Fig.~\ref{recovery_time}. Compared with BBE, the half of data cache space in Sepencr is used to store the C-OTPs. Sepencr thus needs to read half of the data of BBE on recovery. The recovery time of Sepencr is also half of BBE. For a large enough data cache (e.g., 32MB data cache), the recovery time of BBE is less than 0.11s. Moreover, after crashes and reboots, the system needs 10--100s to perform self-check~\cite{huang2021star,selftest}. The recovery time of Sepencr/BBE is negligible.

%% file: discussion.tex
\section{Discussions}
\label{discussion}
\textbf{The difference between dirty and clean data in eADR-based caches.}
An intuitive optimization for an eADR-based NVM system is to distinguish between dirty and clean data in the cache and handle them in different ways~\cite{alshboul2021bbb}. Specifically, the dirty data in cache are modified, and the clean data in cache are not modified. Upon crashes, only the dirty data need to be flushed into NVM. The clean data in cache can be discarded. In the persistent workloads with flush instruction, there are many clean data in cache since the dirty data are actively flushed via the flush instruction during running time. However, due to the simplified flush instructions in eADR-based NVM systems, the data in cache are passively evicted via the cache replacement policy. For example, by using the Least Recently Used (LRU) cache replacement policy, the system prioritizes keeping dirty data in cache and evicting clean data when the dirty data are more likely to be recently used than the clean data, especially in write-intensive workloads. In our experiments, we found that after system warm-up, almost all data in the eADR-based cache are dirty. In our Sepencr/BBE, we also distinguish between clean and dirty data and discard the clean data upon crashes, although there are no performance improvements.

\textbf{The high space overheads in Sepencr.}
In this paper, we abstract three eADR execution models from ideal to practice. In the first two ideal models, eADR supports computation upon crashes, and our design is low-overhead and high-performance. However, the practical Write-Only Model is the only model supported by the current eADR product. Since eADR cannot support the running of the encryption module upon crashes, the design space for ensuring data confidentiality under the Write-Only Model is limited. Our Sepencr incurs high space overheads to store the C-OTPs. However, the evaluation results under multi-core systems demonstrate that the performance of Sepencr is still high even though the cache contention is severe. We leave the optimization of Sepencr in terms of space overheads as our future work.

\textbf{The applicability of the eADR mechanism.}
In this paper, we discuss the encryption scheme in eADR-based NVM systems under different eADR execution models. We also observed that other mechanisms cannot be directly applied to eADR-based NVM systems under the current version of eADR. Specifically, Error Correction Codes (ECCs) computed in the memory controller~\cite{YeHA18} are proposed to enhance data fault tolerance. Upon crashes, the data in the eADR-based cache is directly flushed into NVM without generating ECCs. Moreover, Oblivious Read Access Machine (ORAM) requires complex data processing before flushing data to hide the program access pattern~\cite{che2020multi}. Directly flushing data from cache into NVM in the eADR-based NVM system breaks the principle of ORAM. It is important to consider how to apply the eADR mechanism in different systems.

%% file: related.tex
\section{Related Work}
\label{related}

\textbf{eADR-based NVM systems.} The recently proposed eADR mechanism significantly improves the system performance due to allowing lazy updates and decreasing the number of flush instructions. HTMFS~\cite{yihtmfs} builds a Hardware Transactional Memory in eADR-based NVM systems to achieve both high performance and strong consistency. Dang et al.~\cite{dang2022nvalloc} evaluate their persistent memory allocator in the eADR-based NVM systems. This allocator significantly improves the system performance. BBB~\cite{alshboul2021bbb} provides a battery-backed persist buffer in each core to bridge the gap between the visibility and persistence. BMF~\cite{freij2021bonsai} leverages the cache protected by the backup battery to store the nodes of integrity trees and ensure crash consistency. Horus~\cite{Horus} reduces the number of accessing security metadata upon crashes in eADR-based systems. Unlike these designs, we discuss how to implement encryption in eADR-based NVM systems from the ideal eADR model to the practical eADR model.


\textbf{Data integrity in NVM systems.} To defend against unauthorized modifications, i.e., integrity attacks, the integrity trees are widely used in NVM systems~\cite{GassendSCDD03,naor2006one,koo2018improving,RogersCPS07,awad2019persistently,rakshit2017assure,freij2021bonsai,CostanD16,lei2020efficient,alwadi2021promt}. To reduce the overheads of integrity verification, Janus~\cite{liu2019janus} executes the integrity tree updates with backend operations (e.g., encryption and deduplication) in parallel. Janus also pre-executes the tree updates before the write requests arrive at the memory controller. Freij et al.~\cite{freij2020streamlining} observed that updating the Bonsai Merkle Tree (BMT) with the correct order needs large overheads. They propose the pipelining BMT update scheme to reduce the latency of updating BMT. SCUE ~\cite{huang2021update} skips the modifications of the intermediate nodes to immediately update the root in SIT. Moreover, there exist crash inconsistency problems among tree nodes and user data upon crashes. Anubis~\cite{ZubairA19}, Triad-NVM~\cite{AwadYSNZ19}, STAR~\cite{huang2021star}, and Phoenix~\cite{alwadi2019phoenix} propose different approaches to recovering the integrity trees from crash states with low recovery time. Unlike these designs, our Sepencr and BBE focus on data confidentiality in eADR-based NVM systems. Since BBE and Sepencr still leverage the counters to encrypt data, our designs are orthogonal to these counter-based integrity trees, e.g., BMT and SIT.

%% file: conclusion.tex
\vspace{-0.2cm}
\section{Conclusion}
\label{conclusion}
In order to efficiently bridge the gap between data persistence and confidentiality, this paper comprehensively studies the ideal and practical models of eADR, and proposes BBE and Sepencr in the eADR-based NVM system. BBE under Write-Compute-Order Model leverages the battery of eADR to support the running of the AES engine and XOR gates to encrypt the cached data upon crashes. Moreover, Sepencr under Write-Only Model leverages the outside-the-memory-space addresses to generate the C-OTPs on system-start. These C-OTPs are stored in cache and used to encrypt cached data in case the system crashes. The C-OTPs are only used upon system crashes. During system running time, the data in the system are encrypted by the CME scheme. Experimental results show that our BBE/Sepencr significantly reduces performance overheads compared with intuitive encrypting data in caches.